\documentclass[intlimits,twoside,a4paper]{article}

\usepackage[cp1251]{inputenc}
\usepackage[]{cmpj3}

\usepackage{bm}

\usepackage{color}

\newcommand{\sizeone}{0.97\textwidth} 

\issue{2020}{23}{4}{43706}
\doinumber{10.5488/CMP.23.43706}

\title[Extended Falicov-Kimball model at weak interactions]%
{Extended Falicov-Kimball model at weak onsite and intersite Coulomb interactions}
\author[K.J. Kapcia, J. Krawczyk, R. Lema{\'{n}}ski]{K.J. Kapcia\refaddr{label1},
        J. Krawczyk\refaddr{label2}, R. Lema{\'{n}}ski\refaddr{label2}}
\addresses{
\addr{label1} Institute of Nuclear Physics, Polish Academy of Sciences, ulica W. E. Radzikowskiego 152, PL-31342 Krak\'{o}w, Poland
\addr{label2} Institute of Low Temperature and Structure Research, Polish Academy of Sciences, ulica Ok\'olna 2, PL-50422 Wroc\l{}aw, Poland
}
\date{Received June 10, 2020, in final form August 1, 2020}
\begin{document}

\maketitle

\begin{abstract}
We analyze in detail a behavior of the order parameter  in the half-filled extended Falicov-Kimball model for small Coulomb interactions (both onsite $U$ and intersite $V$). 
The parameter is defined as the difference of localized electron concentrations in both sublattices of the Bethe lattice (in the limit of large coordination number).
Using two methods, namely, the dynamic mean field theory and the Hartree-Fock approximation, we found the ranges of $U$ and $V$ for which the anomalous temperature dependence of the order parameter, characterized by the sharp reduction near $T \approx T_{\textrm{C}}/2$, occurs ($T_{\textrm{C}}$ is the temperature of the continuous order-disorder transition). 
In order to quantitatively describe  this anomaly, we defined a function that measures the departure of the order parameter dependence from the standard mean-field $S=1/2$ Ising-like curve. 
We determined the $U$-dependent critical value $V_{\textrm{C}}$ of $V$ above which the anomaly disappears.
Indicators of the anomalous behavior of the parameter dependence can be also observed  in the temperature dependence of the specific heat.
\end{abstract}

\section{Introduction}
\label{sec:intro}

Correlations in electron systems sometimes cause surprising effects \cite{MicnasRMP1990,GeorgesRMP1996,ImadaRMP1998,FreericksRMP2003,KotliarRmp2006}.
One of the effects is the anomalous behavior of the order parameter $d(\Theta)$ with reduced temperature $\Theta=T/T_{\textrm{c}}$ in the system described by the Falicov-Kimball model (FKM) for small values of the local Coulomb coupling $U$ at the half-filling.
The parameter $d(\Theta)$ is defined as the difference of localized electron concentrations in both sublattices of the alternate lattice (cf. also the equations in section \ref{sec:mfa}).
This anomalous behavior consists in a sudden drop in the value of $d(\Theta)$ near $\Theta=1/2$ and its low, but non-zero, value in the range $1/2<\Theta<1$, before it reaches $0$ at the order-disorder transition temperature at $\Theta = 1$.

The anomalous behavior of $d(\Theta)$ was first noticed by P.G.J. van Dongen and D. Vollhardt, who solved the half-filled FKM exactly on the infinite Bethe lattice in the limit $U \rightarrow 0$ using the dynamical mean field theory (DMFT)~\cite{DongenPRL1990,DongenPRB1992}.
Then, accurate calculations also carried out using the DMFT showed that such a surprising course of $d(\Theta)$ occurs not only in the limiting case $U\rightarrow 0$ but also for small finite $U$ values (roughly for $U<0.01$) for the Bethe lattice as well as for the hyper-cubic lattice in the limit of large coordination number ($z\rightarrow + \infty$; or, equivalently, in the limit of large dimensions $d\rightarrow+\infty$)
\cite{ChenPRB2003,Krawczyk2018}.
Moreover, it was confirmed that, for any finite $U$, function $d(\Theta)$ exhibits the square-root character when $\Theta\rightarrow 1$ \cite{DongenPRB1992,ChenPRB2003,Krawczyk2018}.
A similar anomalous behavior of $d(\Theta)$ was also obtained by means of numerical Monte Carlo calculations for the FKM on the square lattice ($d=2$) \cite{MaskaPRB2006}.

On the other hand, it was reported in reference \cite{DongenPRB1992} that in the extended FKM (EFKM), i.e., after  additional introduction of repulsion between electrons located on adjacent ions with the interaction constant $V$, the shape of the order parameter curve for $V>0$ and $U$ comparable in magnitude takes the standard mean-field $S=1/2$ Ising-like form, i.e., it is completely different from that for $V=0$. 
But again, this outcome was found only in the limiting case of $U\rightarrow 0$. 
Therefore, the question arose: does this anomalous course of the function $d(\Theta)$ occur only at the singular point $V=0$ and disappears at any positive $V$? 
If so, this anomaly could not occur in real systems, because there is always, perhaps even very small, a Coulomb repulsion of electrons located on the neighboring ions.

To clarify this issue, in this work we  systematically study the behavior of $d(\Theta)$  function in a range of small values of $U$ and $V$ couplings. 
Our goal  was to check if there is a finite area in the parameter space $(U, V)$ in which the anomalous course of $d(\Theta)$ occurs, and if so, to determine this area.
Using two methods, namely (i) the DMFT and (ii) the broken-symmetry mean-field Hartree-Fock approximation (HFA), we are able to find $U$-dependent critical value $V_{\textrm{c}}$ of $V$ coupling, which destroys the anomalous behavior of $d(\Theta)$  occurring for $V<V_{\textrm{c}}$.

The work is organized as follows.
In section~\ref{sec:model} we describe the model and the methods used in our analysis.
In section~\ref{sec:orderparameter} the results for $d(\Theta)$ are presented and analyzed by means of the introduced measure of the anomaly of $d(\Theta)$ curve.
Some exemplary temperature dependencies of the specific heat are discussed in section~\ref{sec:specheat}. 
section~\ref{sec:conclusions} includes conclusions and final comments.

\section{Model and methods}
\label{sec:model}

\subsection{Half-filled extended Falicov-Kimball model}
\label{sec:hfefkm}

The Hamiltonian of the EFKM (cf. also references \cite{DongenPRL1990,DongenPRB1992,LemanskiPRB2017,KapciaPRB2019,KapciaPRB2019a,KapciaLemanskiZygmunt}) on a lattice in the second quantization formalism has the following form  
\begin{equation}
\label{eq:ham}
\hat{H}  =  
\frac{t}{\sqrt{z}}\sum_{\left\langle i,j\right\rangle}{ \left( \hat{c}^{\dag}_{i,\downarrow}\hat{c}_{j,\downarrow} + \hat{c}^{\dag}_{j,\downarrow}\hat{c}_{i,\downarrow} \right) }
 + U\sum_i \hat{n}_{i,\uparrow} \hat{n}_{i,\downarrow} 
 +  \frac{2V}{z} \sum_{\left\langle i,j\right\rangle,\sigma,\sigma'} \hat{n}_{i,\sigma}\hat{n}_{j,\sigma'} 
 - \sum_{i,\sigma} \mu_{i,\sigma}\hat{n}_{i,\sigma}, 
\end{equation}
where the denotions used here are the same as those used in references \cite{DongenPRB1992,LemanskiPRB2017,KapciaPRB2019,KapciaPRB2019a,KapciaLemanskiZygmunt}. 
So $\hat{c}^{\dag}_{i,\downarrow}$ ($\hat{c}_{i,\downarrow}$) denotes creation (annihilation) of a fermion (electron) with spin $\sigma$ ($\sigma \in \{ \downarrow, \uparrow \}$) at lattice site $i$ and $\hat{n}_{i,\sigma} = \hat{c}^{\dag}_{i,\downarrow} \hat{c}_{i,\downarrow}$. 
$U$ and $V$ denote onsite and intersite nearest-neighbor, respectively,  density-density Coulomb 
interactions, $z$ is the coordination number,
$\mu_{i,\sigma}$ is the local chemical potential for electrons with spin $\sigma$ at site $i$.
In this work, we consider the case of half-filling, i.e., $\mu_{i,\sigma} = (U + 4V)/2$.
$\left\langle i,j \right\rangle$ denotes the sum over nearest-neighbor pairs and
the prefactors in the first and the third term in equation (\ref{eq:ham}) were chosen so that they yield a finite and non-vanishing contribution to the free energy per site in the limit $z\rightarrow+\infty$.

As one of the simplest models of correlated electron systems, the standard spineless FKM (with $V=0$), as well as some of its extensions have been intensively studied  \cite{KennedyPhysA1986,BrandtMielsch1989,BrandtMielsch1990,BrandtMielsch1991,BrandtUrbanek1992,Watson1995,
FreericksPRB1999,FreericksPRB2000,ZlaticFreericksLemanskiCzycholl2001,WojtkiewiczPRB2001,
LemanskiPRL2002,ShvaikaFreericksPRB2003,GajekJMMM2004,BrydonPRB2005,
FreericksBook2006,Stasyuk2006,HassanPRB2007,MatveevPRB2008,LemanskiPRB2008,Farkasovsky2008,ZondaSSC2009,
YadavEPL2011,Lemanski2014,Farkasovsky2015,Lemanski2016,HamadaJPhysSocJap2017,Farkasovsky2019,
ZondaPRB2019,SmorkaPRB2020,AstleithnerPRB2020}. 
In particular, other than displayed in  (\ref{eq:ham})  extended versions of the FKM include, e.g., those that take into account correlated or extended electron hopping \cite{WojtkiewiczPRB2001,ShvaikaPRB2003}. 
Reviews of the results obtained for the FKM and its various extensions can be found, e.g., in references \cite{Gruber1996,Jedrzejewski2001,FreericksRMP2003,FreericksBook2006,CencarikovaCMP2011}.

In this paper, we analyze the model on the Bethe lattice (Cayley tree) in the limit of large coordination number \cite{Bethe1935,DongenPRB1992,GeorgesRMP1996}.
Originally, the lattice was demonstrated as an infinite connected cycle-free graph with the vertices all having the same coordination number \cite{Bethe1935}.
The statistical mechanics of various lattice models on this graph are usually exactly solvable, due to its very specific topology \cite{OstilliPhysA2012}.
Thus, we use the semi-elliptic density of states (DOS) in the HFA calculations, which is the DOS of non-interacting particles on the Bethe lattice with the coordination number $z\rightarrow \infty$ \cite{GeorgesRMP1996,FreericksRMP2003}. 
The explicit form for  this DOS function (in $z\rightarrow \infty$ limit) is $D_{\textrm{S-E}}(\varepsilon)=(2\piup t^2)^{-1}\sqrt{4t^2-\varepsilon^2}$ for $|\varepsilon| \leqslant 2t $ and $D_{\textrm{S-E}}(\varepsilon) = 0$ for $|\varepsilon | > 2t$, so in all these cases, the half-bandwidth is equal to $2t$. 
However, this DOS can be also treated as an approximation for the DOS for the cubic lattice ($d=3$).
In the rest of the paper, we take $t$ as an energy unit ($t=1$).

\subsection{Mean-field approaches used}
\label{sec:mfa}

In the present work we consider the ordering on the alternate lattice, i.e., the lattice composed of two interpenetrating sublattices.
Thus, one defines two mean-field order parameters (averages) $d$ and $d_1$ as differences between
average occupation of sublattices by  localized and itinerant
particles, respectively (cf. references \cite{DongenPRB1992,Lemanski2014,LemanskiPRB2017,KapciaPRB2019,KapciaPRB2019a,KapciaLemanskiZygmunt}).
Namely,
$d=n^{A}_{\uparrow} - n^{B}_{\uparrow}, \quad d_{1}=n^{A}_{\downarrow}-n^{B}_{\downarrow}$,
where $n^\alpha_\sigma  = \langle \hat{n}_{i,\sigma} \rangle$ for any $i\in \alpha$, where $\alpha = A,B$ denotes
the sublattice index ($n^\alpha_\sigma$ is an average occupation of sublattice $\alpha$ by particles with spin $\sigma$). 
Due to the equivalence of these two sublattices, we restrict ourselves to the investigation of the solutions with $d>0$.
The parameters are associated to charge polarization $\Delta_{\textrm{Q}}$ and staggered magnetization $m_{\textrm{Q}}$ by the following relations: $\Delta_{\textrm{Q}} = (1/2) \left(d+d_1\right)$ and $m_{\textrm{Q}}=(1/2)(d-d_1)$, respectively. 
They are quantities directly determined in experiments usually.

Here, we treat intersite term $V$ within the Hartree-Fock approximation, restricting ourselves to the Hartree terms only, namely, $\hat{n}_{i\sigma}\hat{n}_{j\sigma'}= \hat{n}_{i\sigma} \left\langle \hat{n}_{j\sigma'} \right\rangle + \left\langle \hat{n}_{i\sigma} \right\rangle \hat{n}_{j\sigma'} - \left\langle \hat{n}_{i\sigma}\right\rangle \left\langle \hat{n}_{j\sigma'} \right\rangle$ ($i \neq j$). It is an exact approach for the intersite interaction in the limit of large dimensions \cite{MullerHartmannZPB1989}. 
For the onsite term, we use two complementary methods.

One is the DMFT method, which exactly captures the effects of quantum dynamics due to onsite interaction in the limit of large dimensions \cite{MullerHartmannZPB1989,GeorgesRMP1996,ImadaRMP1998,FreericksRMP2003,KotliarRmp2006}. 
In the case of the FKM, the DMFT solution obtained by using analytical expresions, including those for the total free energy $F$, was presented in references \cite{DongenPRB1992,Lemanski2014,Krawczyk2018}.
Its extention to the case of the EFKM with $V\neq0$ was demonstrated in references \cite{DongenPRB1992,LemanskiPRB2017,KapciaPRB2019,KapciaPRB2019a}, from which we take the equations to obtain the results presented in this work.
The equations are exact in the whole range of $U$ and $V$ parameters and temperature ($z\rightarrow+\infty$  limit), and due to the fact that they do not contain summation over Matsubara frequencies (as it occurs in reference \cite{DongenPRB1992}), they allow one to determine all significant physical quantities practically with any precision even at very low temperatures (see the comment to this in reference \cite{KapciaPRB2019,KapciaPRB2019a}).

The second method is the HFA used to  $U$ term with restriction to only Hartree terms (done similarly as for $V$-term above, but with $j=i$) \cite{MicnasRMP1990,ImadaRMP1998}, which is an approximate method for this model (the treatment of the onsite interaction is not rigorous). 
However, there are some limits in which it gives rigorous results for the EFKM (e.g., in the ground state and for $U=0$), cf. references \cite{LemanskiPRB2017,KapciaLemanskiZygmunt}.   
The equations derived within the HFA for the EFKM are presentend in \cite{LemanskiPRB2017,KapciaLemanskiZygmunt,KapciaJSNM2020}.

\begin{figure}[!t]
\centering
\includegraphics[width=0.99\textwidth]{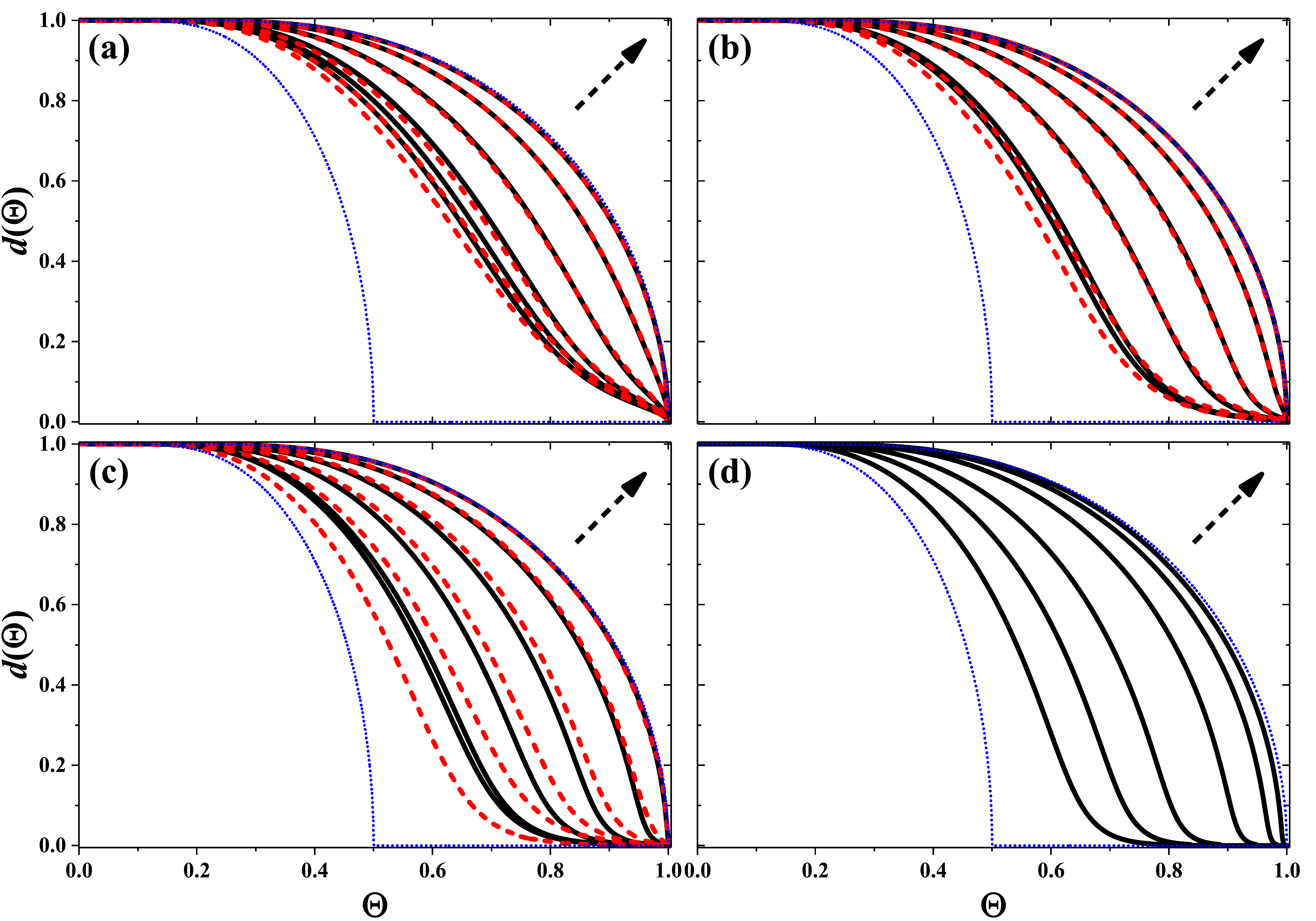}
	\caption{%
	    (Colour online)
		The order parameter $d$ as a function of the reduced temperature $\Theta $ for 
		(a) $U=10^{-2}$ and $V=0$, $10^{-5}$, $3\cdot10^{-5}$, $10^{-4}$, $3\cdot10^{-4}$, $10^{-3}$;
		(b) $U=10^{-3}$ and $V=0$, $10^{-7}$, $10^{-6}$, $3 \cdot 10^{-6}$, $10^{-5}$, $10^{-4}$; 
		(c) $U=10^{-4}$ and $V=0$, $10^{-9}$, $10^{-8}$, $3 \cdot 10^{-8}$, $10^{-7}$, $10^{-6}$; 
		as well as 
		(d) $10^{-6}$ and $V=0$, $10^{-12}$, $3\cdot 10^{-12}$, $10^{-11}$, $3 \cdot 10^{-11}$, $10^{-10}$.
		Mentioned values of $V$ correspond to the curves from the left to the right 
		(the arrows indicate direction of increasing $V$).
		The solid and dashed lines [panels (a)--(c)] correspond to the HFA and the DMFT results, respectively 
		[there is no DMFT results on panel (d)].
		The dotted blue lines denote the standard dependence of the mean-field order parameter 
		$m(T_{\textrm{c}}/2; \Theta)$ and $m(T_{\textrm{c}}; \Theta)$. }
	\label{fig:ordparplot}
\end{figure}

One should also underline that all calculations presented in this work are performed for small and extremely small  interactions as well as temperatures. 
This requires a really high numerical precision, particularly in numerical calculations of derivatives (e.g., specific heat) and integrals [e.g.,  parameter $\xi$ defined in equation (\ref{eq:deviation})].
The results obtained within DMFT have much larger numerical errors than those obtained by the HFA due to the fact that they require minimization of free energy at each temperature~\cite{KapciaPRB2019,KapciaPRB2019a}, whereas the HFA computations are reduced to solving a set of two nonlinear equations \cite{KapciaLemanskiZygmunt}.
Moreover, because the DMFT computations are more time-consuming, every $d(\Theta)$ curve (figure \ref{fig:ordparplot}) obtained within the DMFT and the HFA consists of about $40$ and $5000$ points, respectively.

\section{Results for the order parameter}
\label{sec:orderparameter}

\begin{figure}[t]
\centering
	\includegraphics[width=\sizeone]{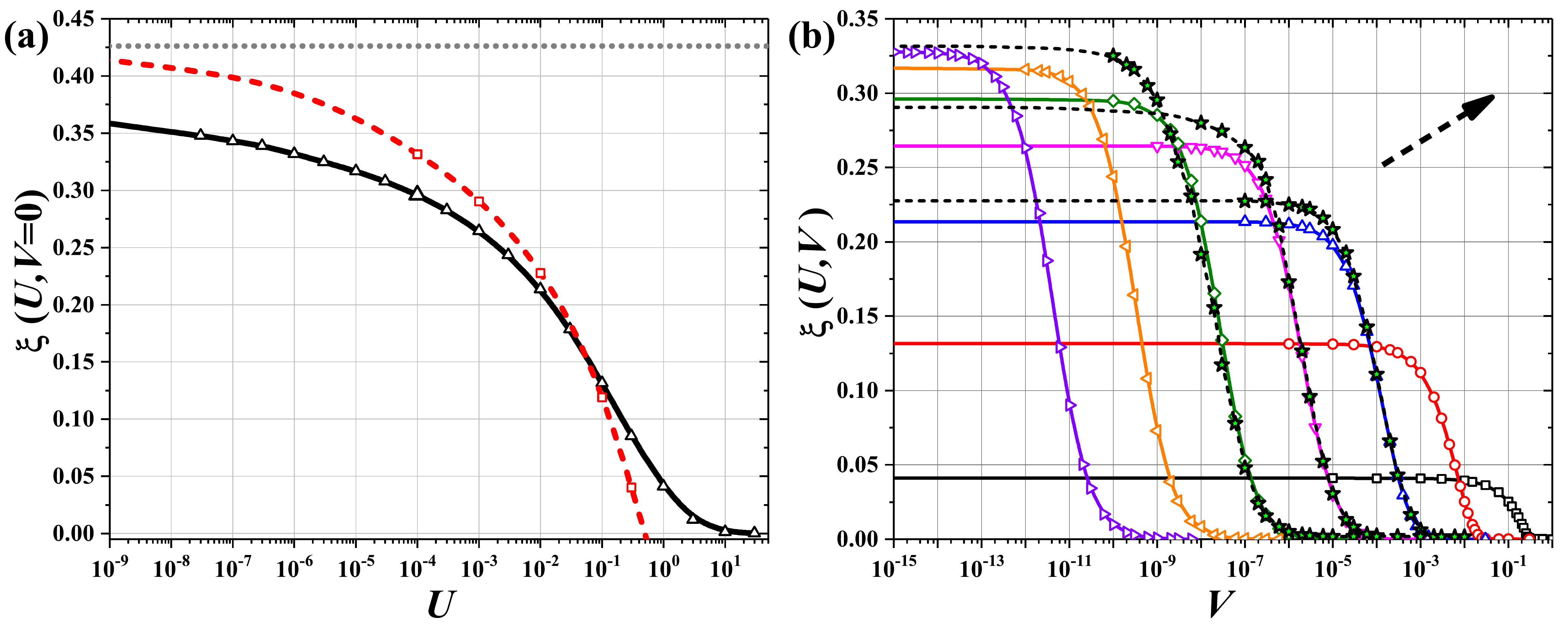}
	\caption{%
	    (Colour online)
		(a)  Deviation $\xi(U,V=0)$ of $d(\Theta)$ curve from the mean-field $S=1/2$ Ising-like form as a function of $U$ for $V=0$.	
		The horizontal dotted line denotes the limiting value of $\xi(U=0,V=0)=0.426$.
		(b) $\xi(U,V)$  as a function of $V$ for different values of 
		$U=1.0$, $10^{-1}$, $10^{-2}$, $10^{-3}$, $10^{-4}$, $10^{-5}$, and $10^{-6}$, 
		from the top right, respectively (the arrows indicate a direction of increasing $U$).
        The DMFT curves are presented for $U=10^{-4}$, $10^{-3}$, and $10^{-2}$ only.
        On both panels, the solid and dashed lines connect the points obtained by the HFA and the DMFT  
        (a guide for eyes), respectively, and the calculated points are denoted by various symbols.	
	}
	\label{fig:ordpardeviations}
\end{figure}

The standard mean-field $S=1/2$ Ising-like temperature-dependence $m(T^{\textrm{MFA}}_{\textrm{C}};\Theta)$ of the order parameter $m$ is described by the solution of the following equation:
\begin{equation}
\label{eq:magnetIsing}
m=\tanh \left( m / \Theta  \right), 
\end{equation}
where $\Theta =T/T^{\textrm{MFA}}_{\textrm{C}}$ and $T^{\textrm{MFA}}_{\textrm{C}}$ is the critical temperature, where $m(T^{\textrm{MFA}}_{\textrm{C}};\Theta)$ tends to $0$ continuously (with simultaneous $\Theta\rightarrow 1$).
Such a dependence can be found, e.g., for the magnetization (staggered magnetization) for the ferromagnetic (antiferromagnetic), respectively, of the Ising model in the absence of the external magnetic field ($B = 0$) by  means of the mean-field approximation, e.g., references \cite{Brush1967,Vives1997}.
For the Hamiltonian of the Ising model defined as $ \hat{H}_{\text{Ising}} = - (J/z) \sum_{\langle i,j \rangle} \hat{S}_{i}\hat{S}_{j} - B \sum_{i} \hat{S}_i$  (where $\hat{S}_i$ is the operator of spin $z$-component at site $i$ with two possible eigenvalues $\hat{S}_i = \pm 1$), the critical temperature within the mean-field approximation is given by $T^{\textrm{MFA}}_{\textrm{C}}=|J|$ (for $B=0$).

Here, we  focus on the dependence of $d(U,V;\Theta)\equiv d(\Theta) $ (where $\Theta=T/T_{\textrm{C}}$ is a reduced temperature) for the EFKM with small $U$ and $V$ interactions. 
In this limit of weak interactions, the model exhibits the second order (continuous) order-disorder transition occurring at the transition temperature $T_{\textrm{C}}$ ($d$ vanishes continously to $0$ at $T_{\textrm{C}}$) \cite{DongenPRB1992,Lemanski2014,KapciaPRB2019,KapciaPRB2019a,KapciaLemanskiZygmunt}.  
Obviously, the HFA overestimates $T_{\textrm{C}}$ and it is always larger than that determined within the DMFT, but for weak $U$ interaction ($U\rightarrow 0$), the difference between them is really small \cite{KapciaLemanskiZygmunt} (for $U,V\rightarrow 0$ also $T_{\textrm{C}}\rightarrow 0$, but for any $|U| + V \neq 0$,  one gets $T_{\textrm{C}}>0$, whereas for $U=0$ and $V=0$,  it is obvious that $T_{\textrm{C}}=0$).  
It was proven that for $U\rightarrow 0$ and $V=0$,  $d(\Theta)$ is a solution $m(T_{\textrm{C}}/2;\Theta)$ of equation (\ref{eq:magnetIsing}) \cite{DongenPRL1990,DongenPRB1992}.
For $V=0$, the $d(\Theta)$ curve exhibits the biggest deviation from $m(T_{\textrm{C}}; \Theta)$ dependence for $U \rightarrow 0$,
where $d(\Theta)=m(T_{\textrm{C}}/2; \Theta)$ \cite{DongenPRB1992,Krawczyk2018,KapciaLemanskiZygmunt} and it decreases to $0$ for $|U| \rightarrow + \infty$, where $d(\Theta)=m(T_{\textrm{C}}; \Theta)$.

It is worth to note that for large $|U|$ or $V$, the dependence $d(\Theta)$ follows the $m(T_{\textrm{C}};\Theta)$ curve (presisely, for $U/V \rightarrow \pm \infty $, $|U| \gg t$, and any $V$; or for $U=0$ and $V \gg t$), but the DMFT predicts $T_{\textrm{C}}\rightarrow 0$ for $U\rightarrow + \infty$ (if $U>2V$) and $T_{\textrm{C}}\rightarrow 2V/k_{\textrm{B}}$ for $U \rightarrow - \infty$ (if $V$ is fixed) \cite{Lemanski2014,KapciaPRB2019,KapciaPRB2019a,MicnasPRB1984,KapciaPhysA2016}, whereas the HFA gives $T_{\textrm{C}}\rightarrow + \infty$ for $U\rightarrow \pm \infty$ (any fixed $V$) \cite{HassanPRB2007,KapciaLemanskiZygmunt}.
For $U=0$ and $V \gg t $,  one gets $T_{\textrm{C}}=V/k_{\textrm{B}}$ in both approaches \cite{KapciaPRB2019,KapciaPRB2019a,MicnasPRB1984,KapciaJSNM2020,KapciaLemanskiZygmunt,KapciaPhysA2016}.
The exact $d(\theta)$ curve obtained within the DMFT can lie even above the  $m(T_{\textrm{C}}; \Theta)$ curve for $U$ larger than $\approx 0.5$, and $d(\theta)$ approaches $m(T_{\textrm{C}}; \Theta)$ from the top (cf. references \cite{ChenPRB2003,Krawczyk2018,KapciaLemanskiZygmunt} for $V=0$). 
Moreover, the HFA results (for $V = 0$) even for $U = 1.0$ exhibit some deviations from the standard
dependence, i.e., mean-field dependence [cf. also figure \ref{fig:ordpardeviations}(a)]. 
However, this large $|U|$ region is out of the scope of the present work.

Several $d(\Theta)$ curves obtained for $U=10^{-2}$,  $U=10^{-3}$,  $U=10^{-4}$, and  $U=10^{-6}$  and for a few values of $V$ are presented in figure \ref{fig:ordparplot}.
They were obtained with the HFA and the DMFT (solid and dashed lines, respectively; for $U=10^{-6}$ only the HFA results are presented).
It is clearly seen that the anomalous behavior present for $U=0$ vanishes with increasing $V$, and for $V$ above some $V_{\textrm{C}}$ the behavior of $d(\Theta)$ curve is almost the same as the standard mean-field $S=1/2$ Ising-like dependence of $m(T_{\textrm{C}}; \Theta)$.
For the largest $V$ presented in the figure, the curves obtained within both approaches (i.e., the HFA and the DMFT) are almost identical.  
For smaller $V$ and $U=10^{-2}$ or  $U=10^{-3}$ [cf. figures \ref{fig:ordparplot}(a) and (b)],  $d(U,V;\Theta)$ dependencies obtained within the DMFT are closer to $m(T_{\textrm{C}}/2; \Theta)$ curve than those obtained with the HFA.
For $U=10^{-4}$ and intermediate $V$, the situation slightly changes  qualitatively and the DMFT curves are above the HFA ones [the DMFT results are closer to $m(T_{\textrm{C}}; \Theta)$], but finally, for $V\rightarrow 0$, $d(\Theta)$ obtained within the DMFT is closer to $m(T_{\textrm{C}}/2; \Theta)$, cf. figure \ref{fig:ordparplot}(c).
It implies that even for such a weak $U$ interaction, the proper inclusion of onsite quantum dynamics (as done in the DMFT) is also important (for really small $V$).
However, it does not mean that the HFA fails for small $U$, because both these approaches give quite comparable results  at finite temperatures, at least qualitatively.   
Let us also note that the function $d(\Theta)$ exhibits the square-root character when $\Theta\rightarrow 1$ for any $|U|+V\neq 0$ (from both the DMFT and the HFA results), as it is expected for the mean-field theories.

\begin{figure}[!t]
\centering
	\includegraphics[width=\sizeone]{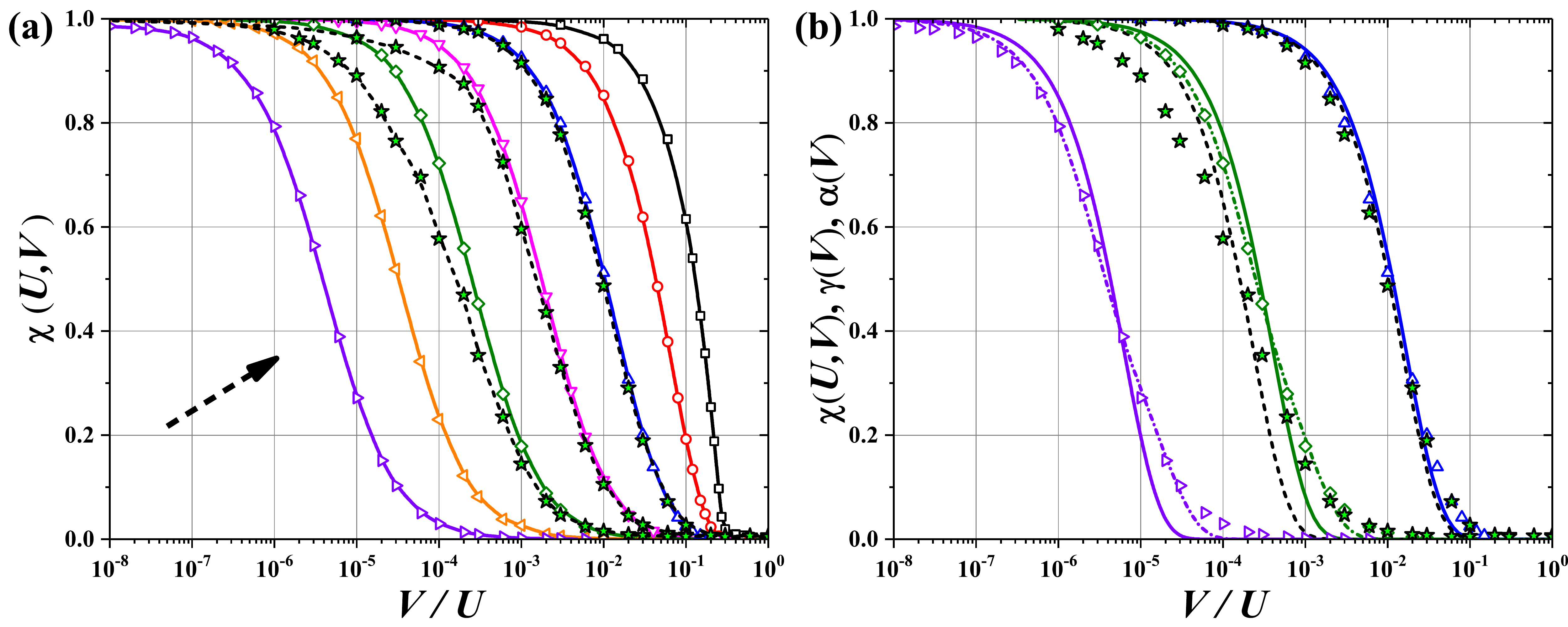}
	\caption{%
	    (Colour online) 
	    (a) Normalized deviation $\chi(U,V)=\xi(U,V)/\xi(U,V=0)$ as a function of $V/U$ for several fixed values of 
		$U$.
		Denotation as in figure \ref{fig:ordpardeviations}(b).
		(b) Exemplary fittings $\gamma(V)$ and $\alpha(V)$  of $\xi(U,V)$ for fixed 
		$U=10^{-6}$, $10^{-4}$, and $10^{-2}$. 
		The solid and dashed lines denote fits $\gamma(V)$ for the HFA and the DMFT 
		(only for $U=10^{-4}$ and $10^{-2}$) data.
		The dash-dotted lines denote fits of $\alpha(V)$ for the HFA data 
		(presented only for $U=10^{-6}$ and $10^{-4}$).
	}
	\label{fig:ordparfits}
\end{figure}

For a quantitative description of the deviation of $d(U,V;\Theta)$ curve from the standard mean-field $S=1/2$ Ising-like dependence of $m(T_{\textrm{C}}; \Theta)$, we introduce the following measure $\xi(U,V)$ of this deviation:
\begin{equation}
\label{eq:deviation}
\xi(U,V) = \int_{0}^{1}  \left[ m(T_{\textrm{C}}; \Theta) - d(U,V;\Theta)  \right] \textrm{d} \Theta.
\end{equation} 
In the investigated range of the model parameters, the integrand function is always positive.
It attains its maximum value for $U,V\rightarrow 0$ [it is equal to $\lim_{U,V\rightarrow 0}\xi(U,V)= \int_{0}^{1}  \left[ m(T_{\textrm{C}}; \Theta) - m(T_{\textrm{C}}/2; \Theta)  \right] \textrm{d} \Theta = 0.426$] and monotonously decreases  with increasing $U$ or $V$ [figure \ref{fig:ordpardeviations}(a)]. 
For small $V$, $\xi(U,V)$ keeps its $\xi(U,V=0)$ value and it is slightly $V$-dependent [cf. figure \ref{fig:ordpardeviations}(b)]. 
At larger $V$, the deviation of $d(U,V;\Theta)$ curve from $m(T_{\textrm{C}},\Theta )$ approaches zero: $\xi(U,V)\approx 0$.
One can identify some $U$-dependent $V_{\textrm{C}}$,  where $\xi(U,V)$ changes from $\xi(U,V=0)$ to $0$.
For $V=0$, the DMFT gives the values of $\xi(U,V=0)$ larger than the HFA. 
Due to the previously mentioned fact that $d(\Theta)$ curve obtained within the DMFT can lie even above $m(T_{\textrm{C}},\Theta)$, the DMFT calculations give $\xi(U,V=0)<0$ for $U$ larger than $U_{\textrm{C}} \approx 0.5$ \cite{ChenPRB2003}, cf. also figure \ref{fig:ordpardeviations}(a).

   \begin{figure}[t]
   \centering
		\includegraphics[width=\sizeone]{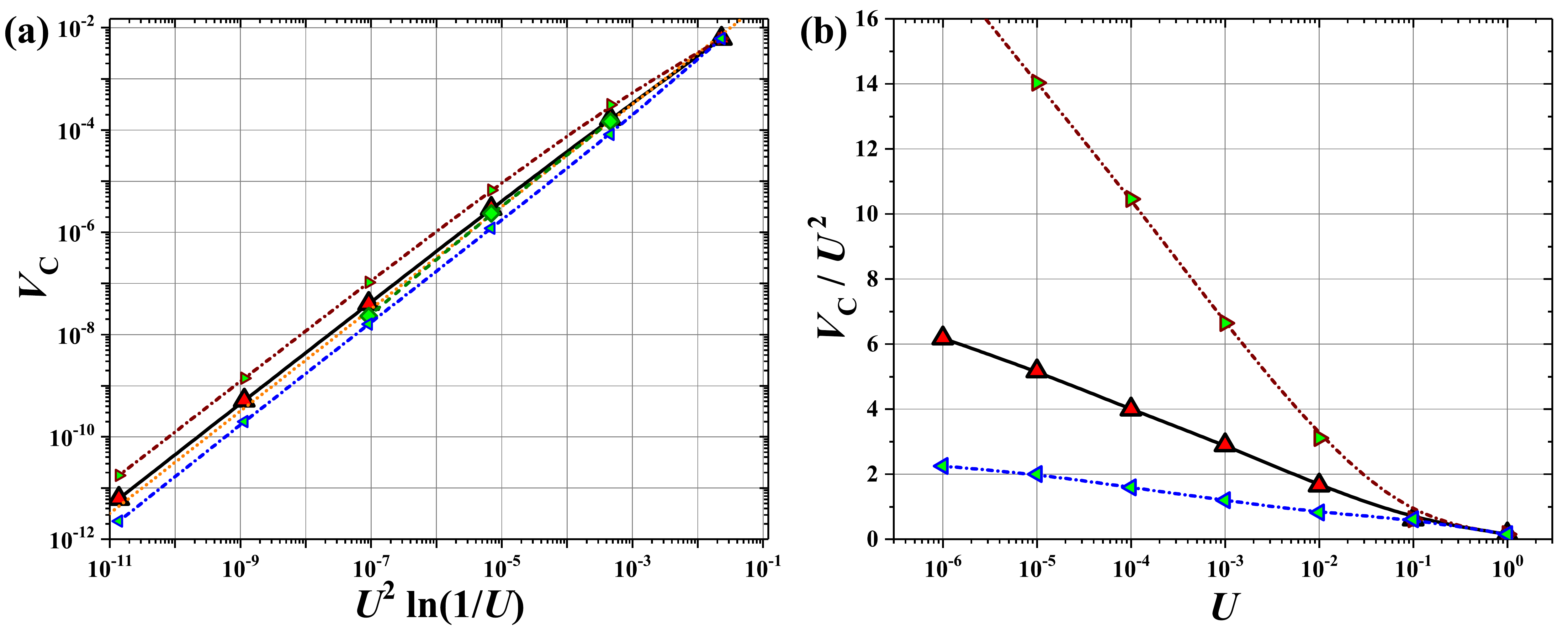}
	\caption{%
		(Colour online)
		The dependence of fitted parameter $V_{\textrm{C}}=\lambda$, $\lambda_1$, 
		and $\lambda_2$ as a function of $U^2\ln(1/U)$  
		(a) and their values normalized by $U^2$  as a function of $U$ (b).
		Dashed and solid lines correspond to $V_{\textrm{C}}$ determined within the DMFT 
		[only panel (a)] and the HFA.
		Dash-dotted	 lines are associated with $\lambda_1$ and $\lambda_2$ obtained from the HFA data.	
		All lines are guide to eyes.
		On panel (a), the dotted line  is a linear function $y=ax$ going through $(0,0)$ with $a=1/\piup$.	
	}
	\label{fig:final}
\end{figure}

More informative quantity to determine $V_{\textrm{C}}$ is $\chi(U,V) =\xi (U,V) /\xi (U,V=0)$ (i.e., $\xi$ normalized by its $V=0$ value).
In figure \ref{fig:ordparfits}(a), the $V/U$-dependence of $\chi(U,V)$ is shown. 
The crossover region of the change of $\chi$ from $1$ to $0$ is clearly noticeable.  
The decay of $\chi$ is slightly faster within the DMFT.    
It turns out that the dependence of $\chi(U,V)$ for fixed $U$ can be quite well described by the single-parameter  exponential-decay function $\gamma(\lambda;V)\equiv \gamma(V)$
\begin{equation}
\label{eq:fittingfunction}
\chi(U,V) \approx \gamma(V) = \exp{\left( - V / \lambda \right)},
\end{equation}  
where $\lambda \equiv V_{\textrm{C}}$ is the $U$-dependent fitting parameter.
This parameter can be treated as an approximation of the effective value of intersite interaction, at which the low-$U$ anomaly of $d(\Theta)$ curve is destroyed: $\gamma(\lambda) = (1/\textrm{e}) \gamma(0)$ ($\textrm{e}\approx 2.71$).
Figure \ref{fig:ordparfits}(b)  presents a few fits $\gamma$ for the data of $\chi$ obtained for several values of $U$.
One see that, indeed, the assumed fitting function $\gamma$ can be considered as an approximation of $V$-dependence of $\chi(U,V)$.  
The dependence of $V_{\textrm{C}}$ as a function of $U$ is shown in figure \ref{fig:final}.
Note that $V_{\textrm{C}}$ is an increasing function of $U$, but $V_{\textrm{C}}/U$ does not exceed $1/2$.
In the ground state, a transition between two ordered phases has been found for $U=2V$, cf. phase diagrams presented in references \cite{LemanskiPRB2017,KapciaPRB2019,KapciaPRB2019a,KapciaLemanskiZygmunt}.
For $U>2V$, the antiferromagnetism dominates over the charge order ($m_{\textrm{Q}}>\Delta_{\textrm{Q}}$ or $d_1<0$), whereas $U<2V$ the charge order is dominant ($m_{\textrm{Q}}<\Delta_{\textrm{Q}}$ or $d_1<0$).
At finite temperatures, due to the exact solution of the model, the location of the boundary is shifted  for $U$ smaller than $2V$ and it is discontinuous~for~any~$U$~and any temperature (with discontinuous changes of $d$ and $d_1$ parameters) \cite{KapciaPRB2019,KapciaPRB2019a}. 
Thus, the anomaly of $d(\Theta)$ dependence is destroyed inside the ordered phase with dominant antiferromagnetic order ($m_{\textrm{Q}}>\Delta_{\textrm{Q}}$).

Obviously, a better description of the $V$-dependence of $\chi$ can be obtained, e.g., by two-parameter  exponential-decay function 
\begin{equation}
\alpha(V) = \tfrac{1}{2}\left[ \exp{\left( -V / \lambda_1 \right)} + \exp{\left( - V / \lambda_2 \right)}\right],
\end{equation} 
particularly, for smaller $U$ [cf. dash-dotted lines in figure \ref{fig:ordparfits}(b)]. 
The fitted parameters $\lambda_1$ and $\lambda_2$ are also presented in figure \ref{fig:final}.

From the data presented in figure \ref{fig:final}, one conclude that the relation between $V_{\textrm{C}}$ and $U$ (for small $U\rightarrow 0$) can be approximated by
\begin{equation}
\label{eq:scalling}
V_{\textrm{C}} \propto U^2 \ln{ \left( U^{-1} \right)}.
\end{equation}
The proportional coefficient in the relation above is estimated to be between $0.25$ and $0.45$ and it  cannot be precisely determined  due to accumulating numerical errors (we are working here in the regime of very small $U$, $V$ and $T_{\textrm{C}}$).
The linear fit to three points obtained for the smallest $U$ gives the value of the coefficient about  $\approx 0.43$. 
One should notice that we have also performed the calculation for the EFKM for $U=0$ (cf. also expresions in reference \cite{KapciaJSNM2020}).

Additionally, we checked that, for $U = 0$ and $V = 10^{-9}$ (the smallest possible value in our calculations), according to equation (\ref{eq:scalling}),  there are no signs of an unusual behavior of $d(\Theta)$ and its dependence almost follows $m(T_{\textrm{C}};\Theta)$ ($\xi\approx 2 \cdot 10^{-4}$ in this case).
For $U=0$ and large $V \gg t$, one can analytically show  that the behavior of $d(\Theta)$ is the same as that of $m(T_{\textrm{C}};\Theta)$.

The FKM exhibits symmetry $U\leftrightarrow-U$, thus the results for $U<0$ are the same as for $U>0$ if $V=0$.
In particular, $d(U=|x|,V=0;\Theta) = d(U=-|x|,V=0;\Theta)$.
This symmetry is destroyed by $V>0$, but for $V\rightarrow 0$, it is obviously restored and the differences between opposite signs of $U$ are negligible.
However, it turns out that $d(\Theta)$ curves for $U>0$ are located above $d(\Theta)$ dependencies obtained for $U<0$ [i.e., $d(U=|x|,V;\Theta) > d(U=-|x|,V;\Theta)$].
Thus, $V_{\textrm{C}}(U=|x|) > V_{\textrm{C}}(U=-|x|)$ and the anomalous behavior  of $d(\Theta)$ curve near $\Theta=1/2$ is destroyed  faster for $U > 0$.
In that sense, the effect of $V$ on the anomalous behavior of $d(\Theta)$  in the phase with dominant charge order (i.e., where $\Delta_{\textrm{Q}}>m_{\textrm{Q}}$) is smaller than that in the phase with dominant antiferromagnetic ordering occurring for $U>2V$ discussed previously.
We do not present the results for $U<0$ in figures due to the fact that the differences are very small (below the thickness of lines in the figures) and they decrease with a decreasing $|U|$.

   \begin{figure}[t]
   \centering
	\includegraphics[width=\sizeone]{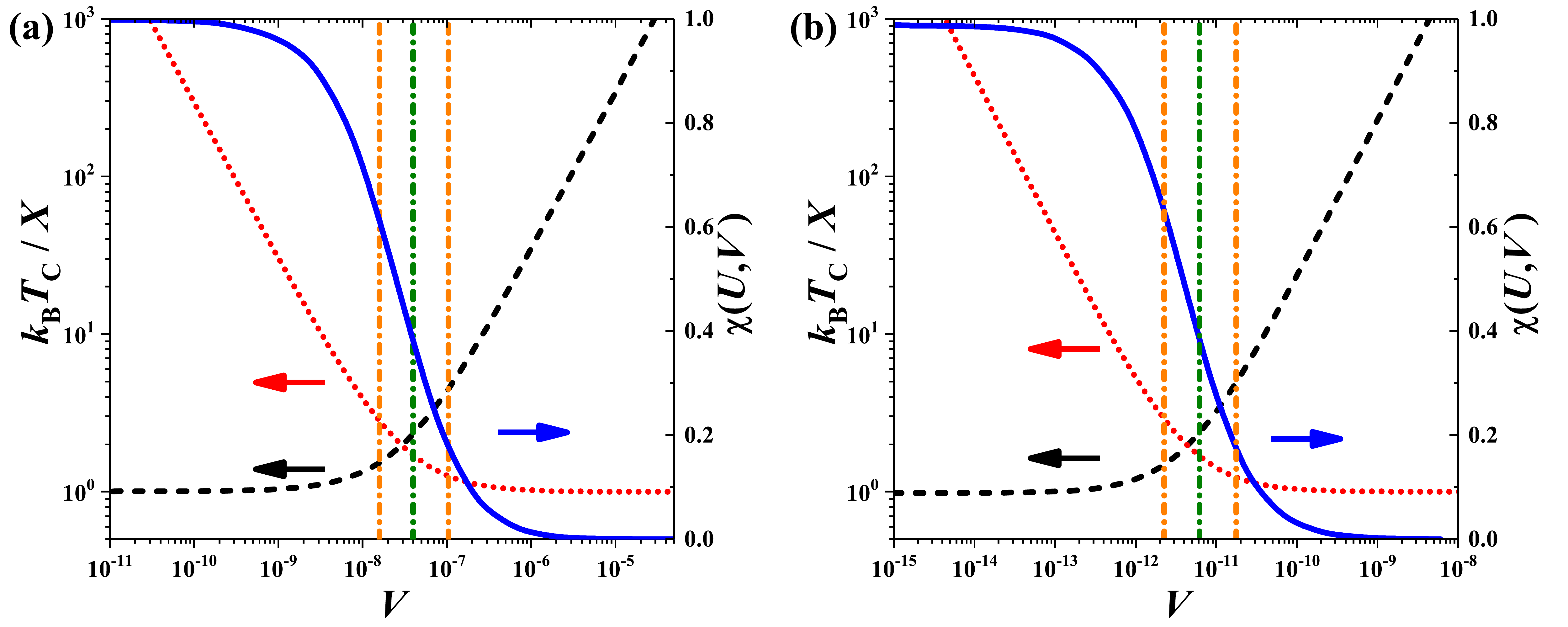}
	\caption{%
	(Colour online)
	$k_{\textrm{B}}T_{\textrm{C}}/X$ [where $X=(2\piup)^{-1}U^{2}\ln(1/U)$ (dashed lines) or $X=V/2$ (dotted lines), the left-hand scale] 
	as a function of $V$ 
	for (a) $U=10^{-4}$ and (b) $U=10^{-6}$. 
	The solid lines correspond to $\chi(U,V)$ dependencies (the right-hand scale).
	Arrows also indicate  the scales for the curves.
	Vertical dash-dotted lines denote values of $\lambda_1$, $\lambda=V_{\textrm{C}}$, and $\lambda_2$ 
	fitting parameters (from the left, respectively).
	The HFA results are only shown.  
	}
	\label{fig:critTc}
\end{figure}

Note that, in the considered range of small $U$ and $V$, one can also distinguish two regions, where $T_{\textrm{C}}$ behaves differently \cite{DongenPRB1992}. 
Namely, for $U \gg V$ ($V\rightarrow 0$), the physics of the model is driven by $U$ interaction and $T_{\textrm{C}} \propto U^2 \ln (1/U)$. 
For larger $V$, where $V$ interaction dominates, $T_{\textrm{C}} \propto V/2$.
These two behaviours are clearly indicated in figure \ref{fig:critTc}.
Here, we plotted $k_{\textrm{B}}T_{\textrm{C}}/X$ ($k_{\textrm{B}}T_{\textrm{C}}$ normalized by $X$), where $X=X_{1}=(2\piup)^{-1}U^{2}\ln(1/U)$ or $X=X_{2}=V/2$. 
These two different regions of $T_{\textrm{C}}$ behaviour are clearly visible.
The first one corresponds to almost constant $k_{\text{B}}T_{\textrm{C}}/X_1$ (linearly decreasing $k_{\text{B}}T_{\textrm{C}}/X_2$), whereas the second one is associated with constant $k_{\text{B}}T_{\textrm{C}}/X_2$ (linear increase of $k_{\text{B}}T_{\textrm{C}}/X_1$).
The crossover region between these two regimes can be characterized by the parameter $V^{*}$ being the solution of $X_1=X_2$, i.e., $V^{*}\approx (U^2/\piup)\ln(1/U)$.       
The interesting observation implied from these plots is that when $\chi$ starts to decrease from its maximal value of $1$, also $k_{\text{B}}T_{\textrm{C}}/X_1$ starts to increase from its minimum value of $1$.
Moreover, when $\chi$ approaches its minimum value of $0$, $k_{\text{B}}T_{\textrm{C}}/X_2$ does not change and takes on its minimum value equal to $1$. 
Thus, there is a connection between the $T_{\textrm{C}}$ and anomaly $\xi$ of $d(\Theta)$ dependence. 
The scaling of $V_{\textrm{C}}$ found here [i.e., equation (\ref{eq:scalling})] is in an agreement with these asymptotic expressions for $T_{\textrm{C}}$.
However, the proportional factor for $V_{\textrm{C}}$ in equation (\ref{eq:scalling}) could be different from that for $V^{*}$.

\section{Specific heat}
\label{sec:specheat}

An anomalous dependence of $d(\Theta)$ also implies  an unusual dependence of the specific heat $c=-T(\partial^2 F/ \partial T^2)$ as a function of reduced temperature $\Theta$.
The results for the FKM with small $U$ were presented in \cite{Krawczyk2018}.
It was found (for $V=0$) that the function $c(\Theta)$ changes its shape from the lambda-type for $U=1$, assuming two maxima in the intermediate range of $0.1 < U < 0.3$, and finally reaching only one maximum around $\Theta=1/2$ when $U<0.01$.

\begin{figure}[t]
   \centering
	\includegraphics[width=\sizeone]{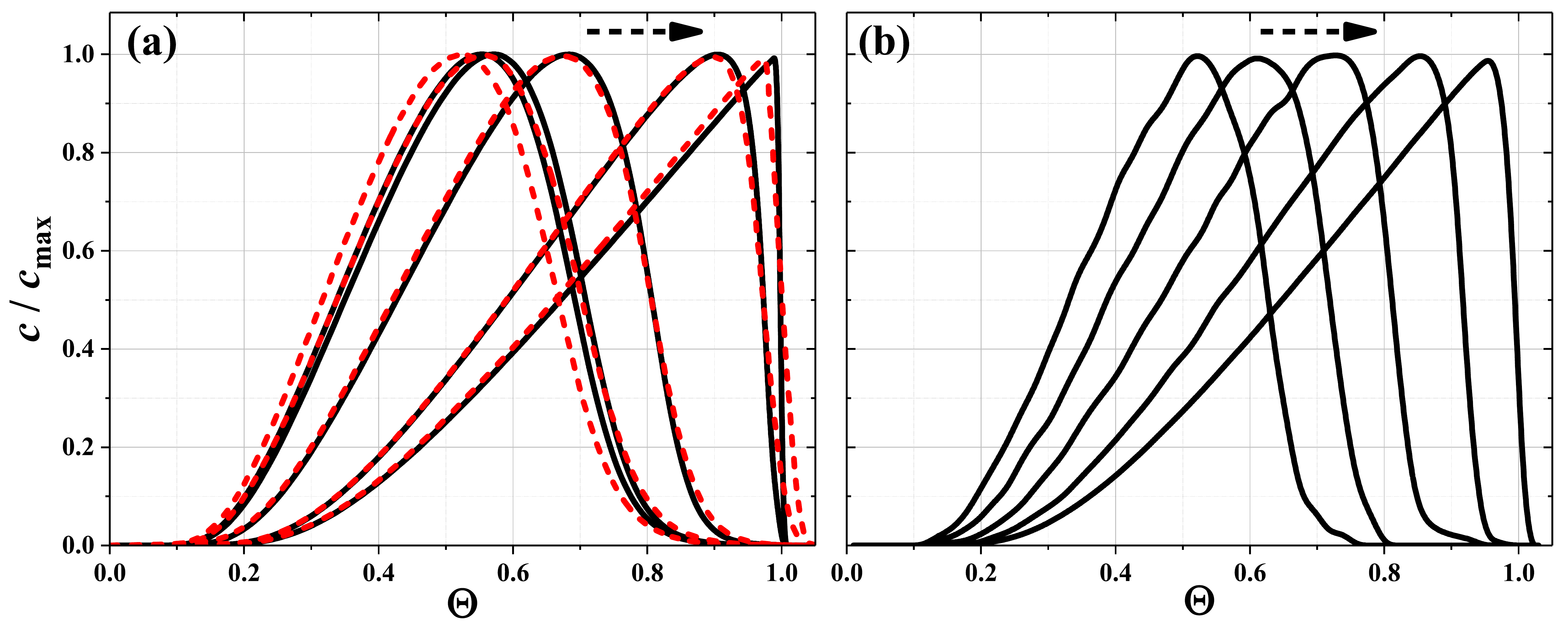}
	\caption{%
		(Colour online)
		The dependence of specific heat $c$ (normalized by its maximum peak value $c_{\textrm{max}}$) 
		as a function of reduced temperature $\Theta$: 
		(a) the results for $U=10^{-3}$ and $V=0$, $10^{-7}$, $10^{-6}$, $10^{-5}$, $10^{-4}$ 
		(solid and dashed lines correspond to the HFA and the DMFT results, respectively) as well as 
		(b) the HFA results for $U=10^{-5}$ and $V=0$, $10^{-10}$, $3\cdot 10^{-10}$, $10^{-9}$, $10^{-8}$.
		The arrows indicate a direction of increasing $V$.
	}
	\label{fig:specificheat}
\end{figure}

A similar evolution from the lambda-type behavior (for  $V>V_{\textrm{C}}$) to a curve with one maximum around $\Theta=1/2$  for $V\rightarrow 0$ (and small $U$) is observed.
Figure \ref{fig:specificheat} presents some exemplary  dependencies of specific heat $c$ (normalized by its maximum value $c_{\textrm{max}}$) for two fixed values of $U$ and increasing values of $V$.
Similarly to the case of $V=0$, the maximum of $c$ occurs somewhat above $\Theta=1/2$, when there is a substantial decrease in the order parameter, although the maxima of $c$ and the derivative of $d$ (with respect to $\Theta$) are sligthly displaced.
With an increasing $V$, it shifts towards higher $\Theta$.
The position of $c_{\textrm{max}}$ calculated within the DMFT is located in slightly lower $\Theta$ than the one obtained within the HFA, which coincides with the presented results for $d(\Theta)$ [cf., e.g., figures \ref{fig:ordparplot}(a) and \ref{fig:specificheat}(a)].
The other drop in the specific heat always also occurs  at the transition point $\Theta=1$, but it is relatively small (for small $V$) in  comparison with $c_{\textrm{max}}$ (due to very small values of $d(\Theta )$ near $T_{\textrm{C}}$) and it is not visible in figure~\ref{fig:specificheat}.
Note that $c$  in the nonordered phase, i.e., for $\Theta>1$, is also small ($c/c_{\textrm{max}}$ is almost $0$) due to small absolute temperatures just above $T_{\textrm{C}}$.
In figure \ref{fig:specificheat}(b) the effects of numerical errors, which accumulate due to twofold derivation and very small changes in both free energy $F$ and $T$, are visible (particularly for $V=0$), although the calculations were performed with a very high precision (in the figure, B-spline curves for the data points are plotted).
This issue is absent for larger $U$.

\section{Conclusions and final remarks}
\label{sec:conclusions}

In this paper we studied the anomalous behavior of the order parameter $d(\Theta )$ versus the reduced temperature 
$\Theta $ in the EFKM for weak couplings $U$ and $V$ using the DMFT and the HFA approaches (in the large coordination number limit on the half-filled Bethe lattice). 
At this regime, both these methods are equivalent, therefore, the results are consistent with each other. 
But since it is easier to perform calculations using the HFA, most of the the data were obtained using this method. We focused mainly on finding the range of Coulomb interaction parameters in which there is an anomalous course of the order parameter as a function of temperature.  
For this purpose, we defined in formula  (\ref{eq:deviation}) the parameter $\xi$ determining the degree of deviation of $d(\Theta )$ from the standard mean-field $S=1/2$ Ising-like curve. This enabled us to determine the degree of this anomaly and to describe its evolution in a quantitative manner for  any small $U$ and $V$ (their ratio is also arbitrary).

The main conclusion from this work is to show that the anomalous behavior of $d(\Theta)$, which is characteristic of the FKM (at $V=0$) at small $U$, also occurs in a certain area of $|U|>0$ and $0<V<V_{\text{cr}}$.  
It also turns out that for a given small $|U|$, the critical value of $V_{\text{cr}}$ is roughly of the same order of magnitude as $U^2$, cf. equation (\ref{eq:scalling}).
Some indicators of the anomalous behavior of $d(\Theta)$ dependence can be also observed in the temperature dependence of the specific heat.

\section*{Acknowledgements}

The authors express their sincere thanks to J.K. Freericks, M.M. Ma\'ska, A.M. Ole\'s, P. Piekarz, and A.~Ptok for very fruitful discussions on some issues raised in this work. 
K.J.K. acknowledges the support from the National Science Centre (NCN, Poland) under Grant SONATINA 1 no. UMO-2017/24/C/ST3/00276.
K.J.K. also appreciates the funding in the frame of a scholarship of the Minister of Science and Higher Education (Poland) for outstanding young scientists (2019 edition, no. 821/STYP/14/2019).
J.K. acknowledges the support from Institute of Low Temperature and Structure Research, Polish Academy of Sciences under the grant for PhD students awarded for the academic year 2018/2019.


\ukrainianpart

\title{Узагальнена модель Фалікова-Кімбала зі слабкою одновузловою та міжвузловою кулонівською взаємодією}
%
\author{К. Я. Капця\refaddr{label1}, Я. Кравчик\refaddr{label2}, Р. Леманський\refaddr{label2}}
\addresses{
\addr{label1} Інститут ядерної фізики, Польська академія наук, вул. В. Е. Радзіковського 152, PL-31342 Краків, Польща
\addr{label2} Інститут низьких температур і досліджень структури, Польська академія наук, вул. Окульна 2, PL-50422 Вроцлав, Польща
}

\makeukrtitle

\begin{abstract}
Ми детально аналізуємо поведінку параметра порядку для узагальненої моделі Фалікова-Кімбала зі слабкою кулонівською взаємодією (одновузловою $U$ і міжвузловою $V$) при половинному заповненні. Параметр порядку визначається як різниця концентрацій локалізованих електронів на різних підґратках ґратки Бете (в границі великих координаційних чисел). За допомогою двох методів, а саме теорії динамічного середнього поля та наближення Хартрі-Фока, ми знайшли інтервали взаємодій $U$ і $V$, при яких спостерігається аномальна температурна залежність параметра порядку, що характеризується його різким зменшенням поблизу $T\approx T_{\textrm{C}}/2$ ($T_{\textrm{C}}$ -- це критична температура неперервного переходу порядок-безлад). Щоб описати цю аномалію кількісно, ми ввели функцію, що визначає відхилення температурної залежності параметра порядку від стандартної кривої Кюрі-Вейса. Ми визначили залежне від $U$ критичне значення міжвузлової взаємодії $V_{\textrm{C}}$, вище якого аномалія зникає. Також, індикатори аномальної поведінки параметра порядку можна спостерігати на температурній залежності питомої теплоти.
\keywords узагальнена модель Фалікова-Кімбала, корельовані електронні системи, ґраткові моделі в конденсованих системах, середньопольові теорії, міжвузлові взаємодії, ґратка Бете
\end{abstract}

%
%
\lastpage
\end{document}